\documentclass[aps,prd,twocolumn,amssymb,eqsecnum,showpacs,showkeyes,secnumarabic,graphics,floatfix,nofootinbib,tightenlines,longbibliography,superscriptaddress]{revtex4-1}

\usepackage[pdftex]{graphicx,color}
\usepackage{bm}
\usepackage{hyperref}
\hypersetup{colorlinks=true,linkcolor=blue,filecolor=magenta,urlcolor=cyan,pdftitle={Pantheon+ homogeneity},pdfpagemode=FullScreen}
\usepackage{amsmath,amssymb}
\usepackage[T1]{fontenc}
\usepackage[utf8]{inputenc}
\usepackage[english]{babel}
\usepackage{lmodern}
\usepackage{longtable}
\usepackage{wasysym}
\usepackage[dvipsnames]{xcolor}
\definecolor{valecol}{rgb}{0,0.5, 1.}

\definecolor{leacol}{rgb}{1.,0.5, 0}

\newcommand{\newc}{\newcommand}
\newc{\D}{\partial}
\newc{\ie}{{\it i.e.} }
\newc{\eg}{{\it e.g.} }
\newc{\etc}{{\it etc.} }
\newc{\etal}{{\it et al.}}
\newc{\lcdm}{$\Lambda$CDM }
\newc{\lcdmnospace}{$\Lambda$CDM}
\newc{\wcdm}{$w$CDM }
\newc{\plcdm}{Planck/$\Lambda$CDM }
\newc{\plcdmnospace}{Planck/$\Lambda$CDM}
\newc{\wlcdm}{WMAP7/$\Lambda$CDM }
\newc{\wlcdmnospace}{WMAP7/$\Lambda$CDM}

\newc{\ra}{\Rightarrow}
\newc{\fs}{$f\sigma_8$}
\newc{\fsz}{$f\sigma_8(z)$}
\newc{\omm}{$\Omega_{0m}$}
\newc{\bea}{\begin{eqnarray*}}
\newc{\eea}{\end{eqnarray*}}
\newc{\be}{\begin{equation}}
\newc{\ee}{\end{equation}}
\newc{\ba}{\begin{eqnarray}}
\newc{\ea}{\end{eqnarray}}

\begin{document}

\title{On the homogeneity of SnIa absolute magnitude in the Pantheon+ sample}

\author{Leandros Perivolaropoulos}\email{leandros@uoi.gr}
\affiliation{Department of Physics, University of Ioannina, GR-45110, Ioannina, Greece}
\author{Foteini Skara}\email{f.skara@uoi.gr}
\affiliation{Department of Physics, University of Ioannina, GR-45110, Ioannina, Greece}

\date{\today}

\begin{abstract}
We have analysed the Pantheon+ sample using a new likelihood model that replaces the single SnIa absolute magnitude parameter $M$ used in the standard likelihood model of Brout et. al.\cite{Brout:2022vxf} with two absolute magnitude parameters $M_<$, $M_>$ and a transition distance $d_{crit}$ that determines the distance at which $M$ changes from $M_<$ to $M_>$. The use of this likelihood dramatically changes the quality of fit to the Pantheon+ sample for a \lcdm background by $\Delta \chi^2=-19.6$ ($\Delta AIC = -15.5$ for two additional parameters). The tension between the  $M_<$ and $M_>$ best fit values is at a level more than $3\sigma$ with a best fit $d_{crit}$ very close to $20Mpc$. The origin of this improvement of fit and $M_<-M_>$ tension is that the new likelihood model, successfully models two signals hidden in the data: 1. The well known systematic effect called {\it volumetric redshift scatter bias} which is due to asymmetric peculiar velocity variations at redshifts $z<0.01$ induced by unequal projected volumes at lower and higher distances compared to a given distance and 2. A mild signal for a change of intrinsic  SnIa luminosity at about $20Mpc$. This interpretation of the results is confirmed 
%using a proper truncation of the Pantheon+ sample as well as Monte Carlo simulations of the observed merged absolute magnitudes $M_i$ of SnIa+Cepheid hosts.
by truncating the $z<0.01$ Hubble diagram data from the Pantheon+ data where the above systematic is dominant and showing that the $M_<-M_>$ tension decreases from above $3\sigma$ to a little less than $2\sigma$. It is also confirmed by performing a Monte Carlo simulation to compare the merged SnIa absolute luminosities $M_i$ of SnIa+Cepheid hosts, obtained from the SH0ES data, with the anticipated luminosities in the context of a homogeneous single absolute magnitude $M$. This simulation shows that the maximum significance of the SnIa luminosity transition ($\Sigma\equiv \frac{|M_>-M_<|}{\sqrt{\sigma_{M_>}^2+\sigma_{M_<}^2}}$) in the real data, is larger than the corresponding maximum significance of $94\%$ of the corresponding homogeneous simulated samples.  
\end{abstract}
\maketitle

\section{Introduction}

The value of the Hubble constant $H_0$ measured by direct local measurements based mainly on  Type Ia supernovae (SnIa) standard candles calibrated using distance ladder methods is at $5\sigma$ tension with the corresponding value of $H_0$ measured indirectly using the sound horizon at last scattering as a standard ruler. This discrepancy constitutes one of the main challenges for the standard cosmological model \lcdm known as the Hubble tension\cite{DiValentino:2021izs,Perivolaropoulos:2021jda,Verde:2019ivm}

The most precise direct method for measuring the Hubble constant is based on observations of SnIa calibrated with Cepheid variable stars in galaxies that host both Cepheid variable stars and SnIa. Cepheids in turn are calibrated using geometric methods (e.g. parallax) in the Milky Way and other nearby anchor galaxies. This is the distance ladder method for the direct measurement of $H_0$\cite{2006hst..prop10802R,Riess:2016jrr,Riess:2020sih,Riess:2021jrx,Freedman_2012}.

Such a distance ladder approach has been implemented recently  by the SH0ES team (Supernovae and H0 for the Equation of State of dark energy) and has lead\cite{Riess:2021jrx} to a best fit value $H_{0}^{R21}=73.04\pm1.04$~km~s$^{-1}$~Mpc$^{-1}$ \cite{Riess:2021jrx}. The corresponding indirect measurement of $H_0$ using the sound horizon at recombination as a standard ruler measured by the CMB perturbations angular power spectrum under the assumption of the validity of the standard cosmological model \lcdm (inverse distance ladder approach) has lead to an even more precise value of $H_0^{P18}=67.36\pm 0.54$~km~s$^{-1}$~Mpc$^{-1}$  \cite{Planck:2018vyg} (see also Refs. \cite{Perivolaropoulos:2021jda,Abdalla:2022yfr,DiValentino:2021izs,Shah:2021onj,Knox:2019rjx,Vagnozzi:2019ezj,Ishak:2018his,Mortsell:2018mfj,Huterer:2017buf,Bernal:2016gxb,Dainotti:2023yrk} for relevant recent reviews). The $5\sigma$ discrepancy (tension) between these two very precise measurements of $H_0$ indicates that most probably at least one of them is not accurate, because the assumptions on which it is based are not valid.

The local direct measurement of SH0ES is consistent with a wide range of other less precise local measurements of $H_0$ using alternative SnIa calibrators \cite{Freedman:2021ahq,Gomez-Valent:2018hwc,Pesce:2020xfe,Freedman:2020dne}, gravitational lensing \cite{Wong:2019kwg,Chen:2019ejq,Birrer:2020tax,Birrer:2018vtm}, gravitational waves \cite{LIGOScientific:2018gmd,Hotokezaka:2018dfi,LIGOScientific:2017adf,DES:2020nay,DES:2019ccw}, gamma-ray bursts as standardizable candles \cite{Cao:2022wlg,Cao:2022yvi,Dainotti:2022rea,Dainotti:2022wli,Dainotti:2013cta}, quasars as distant standard candles \cite{Risaliti:2018reu}, type II supernovae \cite{deJaeger:2022lit,deJaeger:2020zpb}, $\gamma-$ray attenuation \cite{Dominguez:2019jqc} etc. (for recent reviews see Refs. \cite{DiValentino:2021izs, Perivolaropoulos:2021jda}).

The SH0ES measurement relies on the following assumptions:
\begin{itemize}
    \item The measurements of the properties (period, metallicity) and luminosities of Cepheid calibrators and SnIa are accurate and free of unaccounted systematic errors.
    \item The modeling and physical laws involved in the calibration  of Cepheids and SnIa in the three rungs of the distance ladder are accurate and well understood.
\end{itemize}

A recent analysis by the authors \cite{Perivolaropoulos:2022khd} has indicated that a simple variation of the Cepheid/SnIa modeling in the SH0ES analysis introducing a single new degree of freedom can potentially modify the best fit value of $H_0$ in such a way that it may become consistent with the corresponding inverse distance ladder measurement. This new degree of freedom allows for a transition of the SnIa calibrated and corrected intrinsic luminosity (absolute magnitude $M$) at some distance or redshift from a value $M_<$ at low distances (redshifts) to a value $M_>$ at high distances. It would therefore be interesting to introduce this new degree of freedom in the new extended Pantheon+ sample \cite{Brout:2022vxf,Scolnic:2021amr,Brout:2021mpj} which includes many more SnIa than the local SH0ES Cepheid+SnIa sample, to investigate if this degree of freedom is excited by the data.

The Pantheon+ SnIa luminosity sample \cite{Brout:2022vxf,Scolnic:2021amr,Brout:2021mpj}  provides distance moduli derived from 1701 light curves of 1550
SnIa in a redshift range $z\in[0.001,2.26]$ compiled across 18 different surveys. This sample is  significantly improved over the first Pantheon sample of 1048 SnIa\cite{Pan-STARRS1:2017jku}, especially at low redshifts $z$. 

During the past few months when the Pantheon+ sample has been publicly available, a wide range of studies have investigated various aspects of it. In particular, the following aspects of Pantheon+ have been investigated: its consistency with the cosmological principle \cite{Cowell:2022ehf,Sorrenti:2022zat}, the self-consistency level of its covariance \cite{Keeley:2022iba}, its consistency with standard electromagnetism and gravity \cite{Sarracino:2022kve}, the constraints it can provide on modified gravity and generalized dark energy \cite{Brout:2022vxf,Poulin:2022sgp,Wang:2022xdw,Narawade:2022cgb,Bernardo:2022pyz,Wang:2022xdw,Kumar:2023bqj}, the constraints it can provide on early dark energy \cite{Kamionkowski:2022pkx,Simon:2022adh}, the constraints it can provide on the start of cosmic acceleration \cite{Dahiya:2022avg}, the constraints on possible modification of physics at recombination (e.g. electron mass variation) \cite{Lee:2022gzh}, the identification of possible change of the best fit value of $H_0$ when different redshift bins are considered \cite{Jia:2022ycc,Yu:2022wvg,Dainotti:2022bzg,Dainotti:2021pqg}, the effects of binning on its data \cite{Colgain:2022tql,Wang:2022ssr,Pasten:2023rpc}, its consistency with BAO+BBN data \cite{Schoneberg:2022ggi}, the constraints it implies on generalization of the Hubble law \cite{Wang:2022ztz}, constraints on compact object dark matter \cite{Dhawan:2023ekc} etc.

One novel feature of Pantheon+ is that it may be used to infer $H_0$ in addition to cosmological parameters.  This is due to the fact that it includes the distance moduli of SnIa in Cepheid hosts as obtained directly from the distance ladder analysis of SH0ES\cite{Riess:2021jrx}. It also includes the covariance of these SnIa with the SnIa in the Hubble flow. The estimate of $H_0$ was not possible in the first Pantheon sample \cite{Pan-STARRS1:2017jku} because of the degeneracy between $H_0$ and SnIa absolute magnitude $M$. The inclusion of both the apparent magnitude $m_B$ and the distance modulus from Cepheids $\mu_{Ceph}$ for SnIa in Cepheid hosts allows the independent determination of the absolute magnitude $M=m_B-\mu_{Ceph}$ which breaks the degeneracy between $M$ and $H_0$ thus allowing the independent determination of $H_0$ through the Pantheon+ sample.

Due to the new features and data included in the Pantheon+ sample the following questions may be addressed:
\begin{itemize}
    \item Is the best fit value of the SnIa absolute magnitude $M$ consistent among various subsamples of the Pantheon+ sample?
    \item What is the effect of the introduction of new degrees of freedom (e.g. allowing for a change of $M$) on the quality of fit and on the best fit values of $H_0$ and cosmological parameters (e.g. matter density $\Omega_{0m}$)?
\end{itemize}
The goal of the present analysis is to address these questions focusing on the possible inhomogeneities of the standardized/corrected intrinsic luminosity (absolute magnitude) of the SnIa of the Pantheon+ sample as well as possible systematic effects like the  {\it volumetric redshift scatter bias} \cite{Kenworthy:2022jdh,Brout:2022vxf} discussed in Section \ref{III}. Investigations of possible inhomogeneities of other properties of the SnIa (e.g. color or stretch parameters \cite{Wojtak:2022bct}) are also interesting but are beyond the scope of the present study.

The structure of this paper is the following: In the next section \ref{II} we describe the data of the Pantheon+ sample that are relevant for our analysis and describe the method used for the fit of the cosmological parameters, the Hubble parameter $H_0$ and the SnIa absolute magnitude $M$. We then implement this method and obtain the corresponding best fit parameter values for $\Omega_{0m}$, $H_0$ and $M$ in the context of a \lcdm background thus confirming the results of the original analysis of Brout et. al. \cite{Brout:2022vxf} and verifying our implementation of the method described there. In section \ref{III}, we generalize the model and the fitting method by allowing for a transition of the absolute magnitude parameter $M$ at some distance $d_{crit}$ from a value $M_<$ at distances $d<d_{crit}$ to a value $M_>$ at distances $d>d_{crit}$. We find the best fit parameter values for  $\Omega_{0m}$, $H_0$, $M_<$ and $M_>$ with their uncertainties and test the consistency between the best fit values of $M_<$ and $M_>$ and carefully interpret the result taking also into account the volumetric redshift scatter bias. In section \ref{IV} we discuss the statistical properties of the intrinsic luminosities $M_i$ of SnIa in Cepheid hosts as obtained from the SnIa apparent magnitudes $m_{Bi}$ and the Cepheid distance moduli $\mu_{i}^{Ceph}$. Using Monte Carlo simulations and the Kolmogorov-Smirnov (KS) test \cite{kolmogorov_1951}, we check in particular the consistency of the statistical properties of the luminosities among different subsamples of the Pantheon+ \cite{Brout:2022vxf} and SH0ES \cite{Riess:2021jrx}  samples. Finally in Section \ref{V} we review our main results and discuss their interpretation and implications. We also point out possible future extensions of our analysis.

\section{The standard analysis of the Pantheon+ sample for \lcdm}
\label{II}

The Pantheon+ sample is presented through a table (Pantheon+SH0ES.dat) with 1701 rows (plus a header) which includes the data relevant to 1701 SnIa light curves in 47 columns which are described at \href{https://github.com/PantheonPlusSH0ES/DataRelease/tree/main/Pantheon%2B_Data/4_DISTANCES_AND_COVAR}{this url}. 
It also consists of a $1701\times 1701$ covariance matrix $C_{\rm stat+syst}$  which represents the covariance between SnIa  due to systematic and statistical distance moluli uncertainties as described below. The relevant columns for our analysis are the following:
\begin{itemize}
\item {\bf Column 3:} Hubble Diagram Redshift (with CMB and peculiar velocity corrections).
\item {\bf Columns 9-10:} $m_B$ corrected/standardized SnIa apparent magnitude and its uncertainty as obtained from the diagonal of the covariance matrix which also includes peculiar velocity induced, redshift uncertainties.
\item {\bf Columns 11-12:} $\mu=m_B-M_{SH0ES}$ corrected/standardized distance moduli where the absolute SnIa magnitude $M_{SH0ES}=-19.253$ has been determined from SH0ES Cepheid host distances \cite{Riess:2021jrx}. Its uncertainty as obtained from the diagonal of the covariance matrix is included in column 12. Column 11 is superfluous as it is trivially obtained from column 9 by subtracting $M_{SH0ES}$.
\item {\bf Column 13:}   $\mu_{Ceph}$ corrected/standardized distance moduli of the SnIa hosts as obtained from the SH0ES distance ladder analysis \cite{Riess:2021jrx} in the context of the $H_0$ distance ladder measurement. The uncertainty of  $\mu_{Ceph}$ is not included in this Table but it is incorporated in the covariance matrix. This column has entries only in the rows which correspond to SnIa in Cepheid hosts. The rest of the rows have an entry '-9' in this column.
\item {\bf Column 14:}  Takes the value 1 if the SnIa of the row is in Cepheid host and 0 otherwise.
\end{itemize}
In this section we follow \cite{Brout:2022vxf} and use the above described Pantheon+ data to constrain the Hubble parameter $H_0=100\,h$~km~s$^{-1}$~Mpc$^{-1}$, the SnIa absolute magnitude $M$ and the matter density parameter $\Omega_{0m}$ by minimizing a $\chi^2$ likelihood:
\begin{equation}
\label{eq:chi2a}
%-chisq - ln(det(cov)) / 2
\chi^2 = \vec{Q}^T\cdot(C_{\rm stat+syst})^{-1}\cdot\vec{Q} ,
\end{equation}
where $\vec{Q}$ is a vector with dimension 1701 and components which are usually defined as 
\begin{equation}
\label{eq:qdef} Q_i = m_{Bi}-M - \mu_{{\rm model}}(z_i) ,
\end{equation}
where $m_{Bi}-M=\mu_i$  is the distance molulus of the $i^{th}$ SnIa and $\mu_{{\rm model}}(z_i)$ is the corresponding distance modulus as predicted by the assumed background cosmological model parametrization which in the present analysis is assumed to be \lcdm. Thus we have 
\begin{equation}
\mu_{{\rm model}}(z_i) = 5\log(d_L(z_i)/Mpc)+25,
\end{equation} 
where the luminosity distance $d_L(z)$ is 
\begin{equation}
\label{eq:dl}
d_L(z) = (1+z)c\int_0^{z}\frac{dz^\prime}{H(z^\prime)},
\end{equation}
where $c$ is the speed of light and in a \lcdm background
\begin{equation}
H(z) = {H_0}\ \sqrt{\Omega_M(1+z)^3+\Omega_{\Lambda}}.
\label{eq:hz}
\end{equation}
The parameters $M$ and $H_0$ appear in Eqs. (\ref{eq:chi2a}), (\ref{eq:qdef}) only through the combination ${\cal M}\equiv M-5 log(H_0\cdot Mpc/c)$ and therefore they are degenerate and can not be estimated separately. In order to break this degeneracy, $M$ can be estimated separately using the distance ladder approach by calibrating SnIa using Cepheids as was done with the previous Pantheon sample\cite{Scolnic:2017caz}. 

In the Pantheon+ sample this degeneracy is broken within the analysis by modifying the likelihood model (\ref{eq:qdef}) to include the distance moduli of SnIa in Cepheid hosts which can constrain $M$ independently. Thus the vector $\vec Q$ in the likelihood definition (\ref{eq:qdef}) is modified as follows \cite{Brout:2022vxf}
\begin{equation}
\label{eq:qprimedef}
 Q^\prime_i=
        \begin{cases}
            m_{Bi}-M - \mu_i^{{\rm Ceph}} & i \in \text{Cepheid hosts} \\
            m_{Bi}-M - \mu_{{\rm model}}(z_i) &\text{otherwise} ,
        \end{cases}
\end{equation}
where $\mu_i^{Ceph}$ is the distance modulus of the Cepheid host of the $i^{th}$ SnIa which is measured independently in the context of the distance ladder with Cepheid calibrators \cite{Riess:2021jrx}. The novel feature of Pantheon+ is that the components $Q'_i$ that correspond to SnIa in Cepheid hosts are now fully incorporated in the sample and correlated with the rest of the SnIa through the provided covariance matrix. Thus, the degeneracy between $M$ and $H_0$ is broken and the three parameters $M$, $H_0$ and $\Omega_{0m}$ can be fit  in the context of a \lcdm background by minimizing 
\begin{equation}
\label{eq:chi2b}
\chi'^2(M,H_0,\Omega_{0m}) = \vec{Q^\prime}^T\cdot(C_{\rm stat+syst})^{-1}\cdot\vec{Q^\prime} ,
\end{equation}
where $C_{\rm stat+syst}$ denotes the covariance matrix provided with the Pantheon+ data including both statistical and systematic uncertainties. 
\begin{figure*}
\centering
\includegraphics[width = 0.98\textwidth]{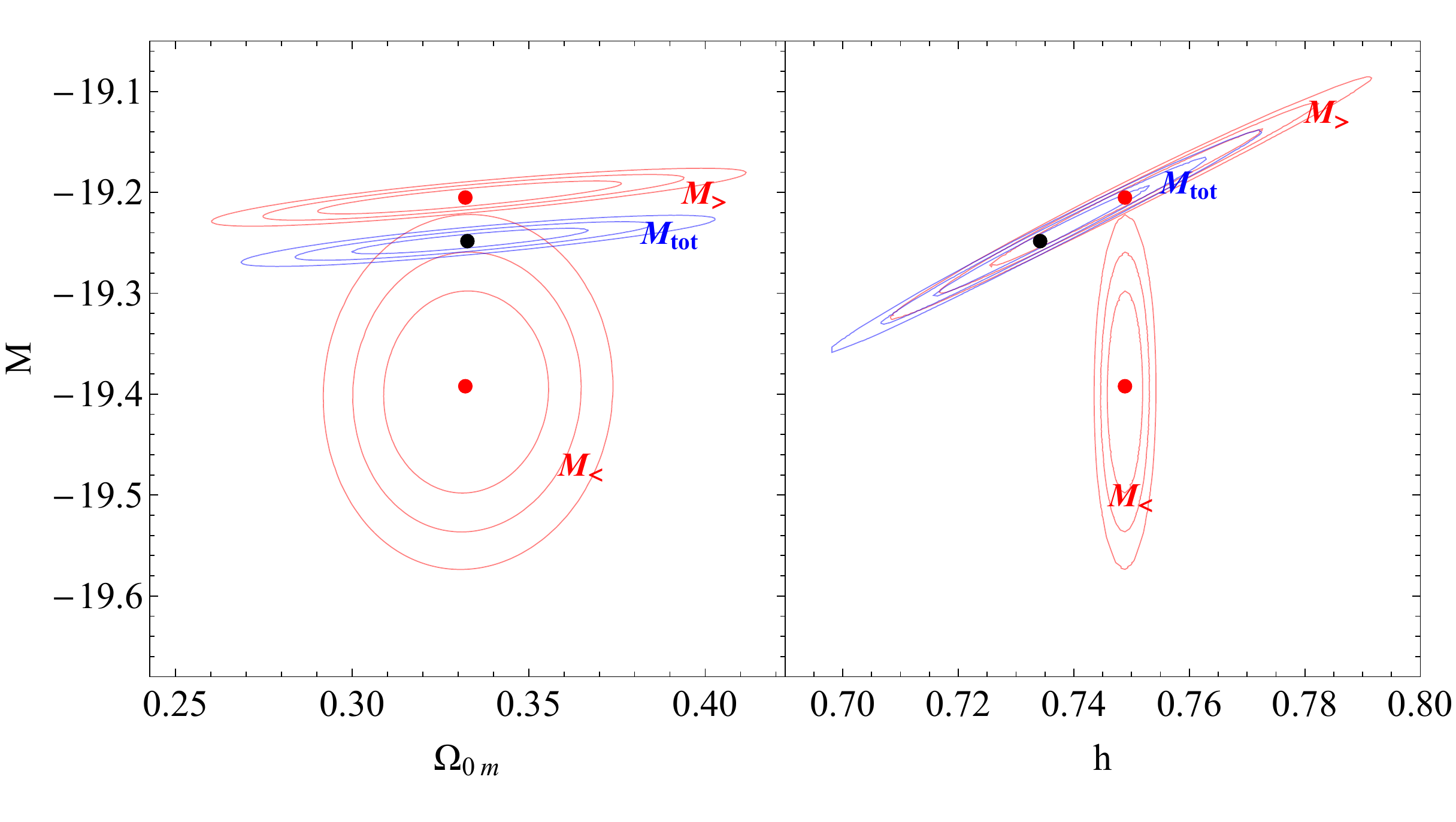}
\caption{Blue contours: The $1-3\sigma$ likelihood contours for the parameters $M$, $h$ and \omm in the context of a \lcdm background using the standard likelihood (\ref{eq:qprimedef}). Red contours: The $1-3\sigma$ likelihood contours for the parameters $M_<$, $M_>$, $h$ and $\Omega_{0m}$  for a \lcdm background in the context of the new likelihood model (\ref{eq:qdprime}).}
\label{fig1}
\end{figure*} 
We have obtained the best fit parameter values for $M$, $H_0$ and $\Omega_{0m}$ and constructed the $1\sigma-3\sigma$ likelihood contours by minimizing $\chi'^2$ of Eq. (\ref{eq:chi2b}) using a simple Mathematica v12 code which is publicly available. 

\begin{figure*}
\centering
\includegraphics[width = 0.98\textwidth]{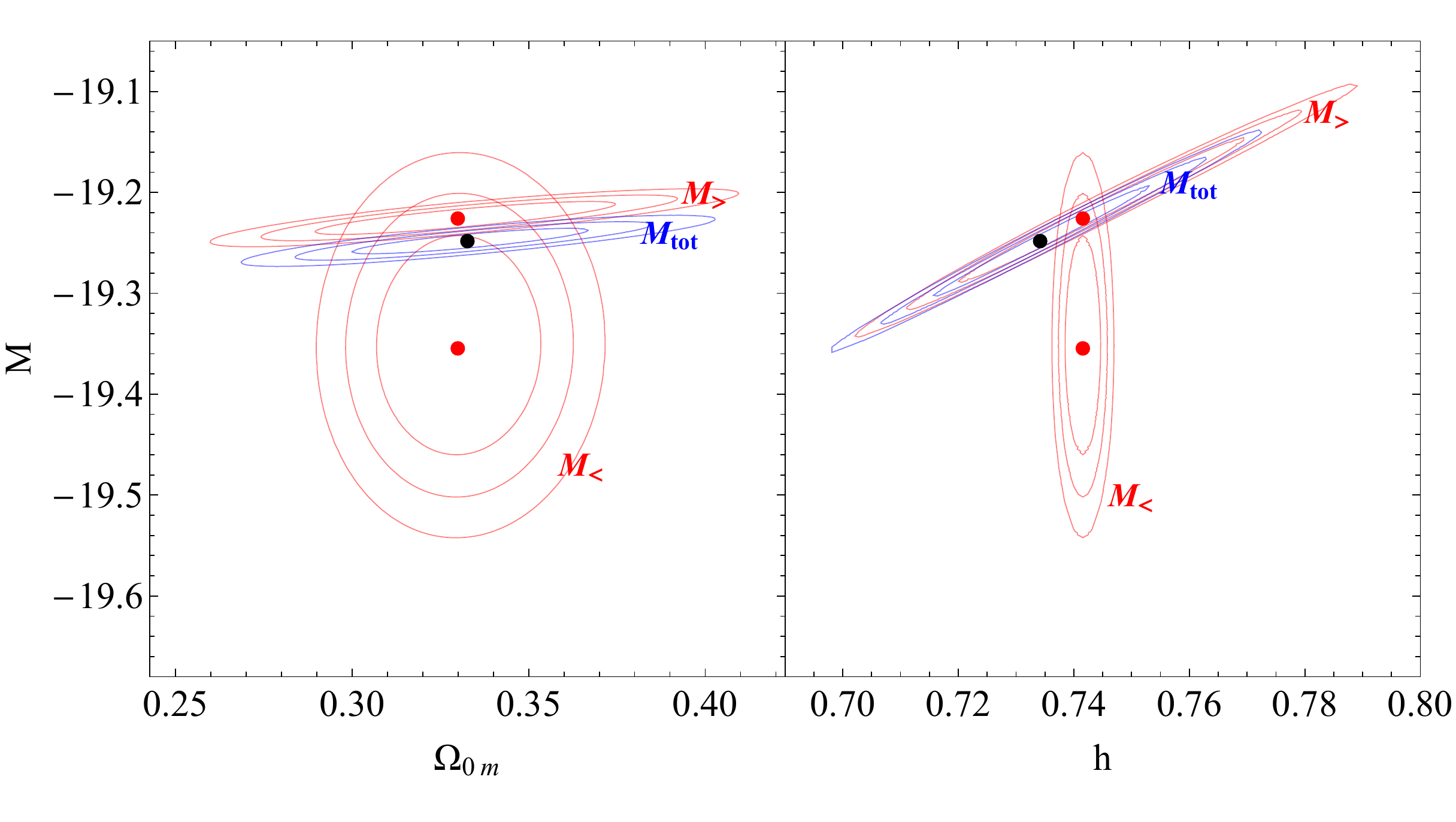}
\caption{Same as Fig. \ref{fig1} but using the likelihood model (\ref{eq:qtprime}) which removes the Hubble diagram data with $z<0.01$ to avoid the volumetric redshift scatter bias which tends to amplify possible intrinsic SnIa luminosity inhomogeneities in the data. As expected, the statistical significance of the apparent SnIa inhomogeneity has been significantly reduced but it has not disappeared. }
\label{fig2}
\end{figure*} 

The uncertainties for each one of the three best fit parameters were obtained using the square roots of the diagonal elements of the parameter covariance matrix which is the inverse of the Fisher matrix defined as
\be
F_{ij}=\frac{1}{2}\frac{\partial^2 \chi'^2(p_1,p_2,p_3)}{\partial p_i \partial p_j},
\label{fisherdef}
\ee
where $i,j=1,2,3$ and the parameters $p_1,p_2,p_3$ correspond to $M$, $h$ and $\Omega_{0m}$. We thus find
\ba
M =&-19.25\pm 0.03 ,\\
h =& 0.734 \pm 0.01 , \label{hstandlik}\\
\Omega_{0m} =& 0.333\pm 0.018 , \label{omstand}
\ea
which is in excellent agreement with the best fit values for $h$ and \omm  $ $ reported in \cite{Brout:2022vxf}. The corresponding parameter likelihood contours are the blue contours shown in Figs. \ref{fig1}, \ref{fig2}.

At the minimum we also find $\chi'^2_{min}=1522.98$ which corresponds to a $\chi'^2$ per degree of freedom of about $0.9$. This is less than 1 and may indicate a possible overestimation of the uncertainties in the covariance matrix as pointed out recently in Ref. \cite{Keeley:2022iba}.

\section{Generalized analysis allowing transition of SnIa luminosity.}
\label{III}
In order to test the homogeneity of the SnIa absolute magnitude parameter $M$, we now generalize the model of the previous section not by allowing more cosmological parameters but by allowing a change of the absolute magnitude at a distance $d_{crit}$ such that it takes the form
\begin{equation}
\label{eq:mlms}
M=
        \begin{cases}
            M_< & d<d_{crit} \\
             M_> & d>d_{crit} ,
        \end{cases}
\end{equation}
The magnitude transition critical distance $d_{crit}$ may be associated with a critical distance modulus through the relation $\mu_{crit}=5 log(d_{crit}/Mpc)+25$. By introducing this degree of freedom in $\chi'^2$ we obtain a generalized $\chi''^2(M_<,M_>,h,\Omega_{0m},d_{crit})$ defined by using a vector $\vec{Q}''$ of the form
\begin{widetext}
\begin{equation}
\label{eq:qdprime}
 Q''_i=
        \begin{cases}
            m_{Bi}-M_< - \mu_i^{{\rm Cepheid}} & {\rm iff} \; \mu_{i,S}<\mu_{crit},{\rm and}\; i \in \text{Cepheid hosts} \\
            m_{Bi}-M_> - \mu_i^{{\rm Cepheid}} & {\rm iff} \; \mu_{i,S}>\mu_{crit},{\rm and}\; i \in \text{Cepheid hosts}\\
            m_{Bi}-M_< - \mu_{{\rm model}}(z_i) & {\rm iff} \; \mu_{i,S}<\mu_{crit},{\rm and}\; i \notin \text{Cepheid hosts}\\ 
            m_{Bi}-M_> - \mu_{{\rm model}}(z_i) & {\rm iff} \; \mu_{i,S}>\mu_{crit},{\rm and}\; i \notin \text{Cepheid hosts},
        \end{cases}
\end{equation}
\end{widetext}
in the expression (\ref{eq:chi2a}) for $\chi^2$. Here $\mu_{i,S}\equiv m_B - M_{SH0ES}\equiv m_B + 19.253$. Minimizing 
\begin{equation}
\label{eq:chi2c}
\chi''^2(M_<,M_>,H_0,\Omega_{0m},d_{crit}) = \vec{Q''}^T\cdot(C_{\rm stat+syst})^{-1}\cdot\vec{Q''} ,
\end{equation}
with respect to the five indicated parameters we find the following best fit parameter values
\ba
M_< =&-19.392\pm 0.05,\\
M_> =&-19.205\pm 0.03, \\
h =& 0.749 \pm 0.01, \label{hnlm} \\
\Omega_{0m} =& 0.332\pm 0.02, \label{omnlm}\\
d_{crit} =& 19.95\pm 0.1 Mpc,
\ea
with $\chi_{min}^2 = 1503.38$. The corresponding likelihood contours are shown in Fig. \ref{fig1}. The quality of fit is significantly improved compared to the standard likelihood (\ref{eq:chi2b}) with $\Delta \chi^2=-19.6$ compared to the baseline model of \cite{Brout:2022vxf}  which has $\chi^2 = 1522.98$. Thus we have a reduction of the  Akaike Information Criterion (AIC) by $\Delta AIC=-15.5$ for the model with 2 additional parameters $d_{crit}$ and $M_<$ which is much larger than 6 and implies strong evidence for the transition model.  Therefore the likelihood model (\ref{eq:qdprime})  fits the Pantheon+ data much more efficiently than the standard likelihood (\ref{eq:qprimedef}) used by Brout et. al. \cite{Brout:2022vxf}. As discussed below, there are two reasons for this improvement of the quality of fit: {\it the volumetric redshift scatter bias} and a mild evidence for an intrinsic transition of the SnIa absolute luminosity.

A well known systematic that may create an apparent inhomogeneity in the low redshift ($z<0.01$) Hubble diagram data of the Pantheon+ sample is the {\it volumetric redshift scatter bias} \cite{Kenworthy:2022jdh,Brout:2022vxf}.  This is an asymmetric velocity variation of SnIa hosts between volumes at higher and lower distances compared to a given distance corresponding to a Hubble diagram redshift.
A given velocity variation magnitude projects more galaxies to a given distance from a higher distance than from a lower distance because the higher distance volume is larger. Thus, there are more galaxies at higher redshifts (where the volume is larger) projected to the correct Hubble diagram redshift due to a given peculiar velocity variation, compared to galaxies at lower redshifts where the volume is smaller. Due to this asymmetry, the Hubble diagram distance projected to a given redshift is higher than the real distance. This effect biases the measured distance modulus $\mu(z)$  to higher values (see eg the low $z$ part of Fig. 4 of \cite{Brout:2022vxf}) creating the impression of a lower SnIa absolute magnitude $M=m_B(z)-\mu(z)$ at $z<0.01$ where the peculiar velocity noise compared to the Hubble flow is more important. 

One way to deal with this bias is to fit for it with additional parameters in the likelihood model (eg $M_<$). In this case it may not be possible to distinguish this bias from a true physical variation of the SnIa intrinsic luminosity but it will significantly improve the quality of fit to the Pantheon+ data of all cosmological models.

An alternative approach would be to eliminate the effects of this bias (along with potentially useful information) by excluding Hubble diagram redshift datapoints at $z<0.01$ from the fit of the low/high distance SnIa absolute magnitudes $M_<$, $M_>$. This may be achieved by the following likelihood model which removes all Hubble diagram data with $z<0.01$:

\begin{widetext}
\begin{equation}
\label{eq:qtprime}
 Q'''_i=
        \begin{cases}
            m_{Bi}-M_< - \mu_i^{{\rm Cepheid}} & {\rm iff} \; \mu_{i,S}<\mu_{crit},{\rm and}\; i \in \text{Cepheid hosts} \\
            m_{Bi}-M_> - \mu_i^{{\rm Cepheid}} & {\rm iff} \; \mu_{i,S}>\mu_{crit},{\rm and}\; i \in \text{Cepheid hosts}\\
             0 & {\rm iff} \; z_i<0.01\\
            m_{Bi}-M_< - \mu_{{\rm model}}(z_i) & {\rm iff} \; z_i>0.01 \; {\rm and}\; \mu_{i,S}<\mu_{crit},{\rm and}\; i \notin \text{Cepheid hosts}\\ 
            m_{Bi}-M_> - \mu_{{\rm model}}(z_i) & {\rm iff} z_i>0.01 \; {\rm and}\;\; \mu_{i,S}>\mu_{crit},{\rm and}\; i \notin \text{Cepheid hosts},
        \end{cases}
\end{equation}
\end{widetext}
Thus, minimizing
\begin{equation}
\label{eq:chi2d}
\chi'''^2(M_<,M_>,H_0,\Omega_{0m},d_{crit}) = \vec{Q'''}^T\cdot(C_{\rm stat+syst})^{-1}\cdot\vec{Q'''} ,
\end{equation}
leads to the best fit parameter values
\ba
M_< =&-19.355\pm 0.05,\\
M_> =&-19.226\pm 0.03, \\
h =& 0.74 \pm 0.01, \\
\Omega_{0m} =& 0.33\pm 0.02, \\
d_{crit} =& 19.95\pm 0.1 Mpc,
\ea
The contours for this likelihood model are shown in Fig. \ref{fig2} and we find $\chi_{min}^2=1445.7$ which is  $\Delta \chi^2 = -7.5$ lower compared to the corresponding  standard likelihood analysis (\ref{eq:chi2b})  with the $z<0.01$ datapoints removed.  Thus we have a reduction of the  Akaike Information Criterion (AIC) by $\Delta AIC=-3.5$ for the model (\ref{eq:qtprime}) with 2 additional parameters $d_{crit}$ and $M_<$ which is less than 6 but larger than 2 and implies mild evidence for the transition model even after the $z<0.01$ datapoints are removed.

Based on the above new likelihood models, it becomes clear that at $d_{crit}\simeq 20Mpc$ there is a discrepancy between the best fit values parameter values $M_<$ and $M_>$ at a $3\sigma$ level. The discrepancy between these two parameters is partly due to the volumetric redshift scatter bias (which is not related to SnIa intrinsic luminosities) and partly due mild but persisting data hints for a transition of the  SnIa intrinsic luminosities.  

The modeling of this discrepancy with the introduced new degree of freedom induces no change in the best fit value of \omm.  However, some  change is observed in the best fit value of $h$ within its $1-2\sigma$ range (compare Eqs. (\ref{hstandlik}) and (\ref{hnlm})). 

The question that raised therefore is the following: {\it Would other dynamical dark energy parameters be affected by the modeling of this discrepancy with new degrees of freedom?} The answer of this question is beyond the scope of the present analysis but if it is positive for some dynamical dark energy models, then the new likelihood model (\ref{eq:qdprime}) which provides an overall much better quality of fit, may be preferable over the standard Pantheon+ likelihood model (\ref{eq:qprimedef}) for cosmological model fits.

%In the SH0ES data analysis for the same value of $d_{crit}\simeq 20Mpc$,  the same degree of freedom induced no significant shift in the best fit value of $h$ \cite{Perivolaropoulos:2022khd} even though hints of discrepancy between $M_<$ and $M_>$ were observed at similar $d_{crit}$ albeit at lower statistical significance (see e.g. Figs. 8, 9 of Ref. \cite{Perivolaropoulos:2022khd}).

The best fit value $M_<$ of the low distance SnIa absolute magnitude is fully consistent with the inverse distance ladder best fit value $M=-19.4\pm 0.027$ \cite{Marra:2021fvf} while the Hubble parameter best fit is about $1.5\sigma$ higher than the corresponding best fit value in the context of the standard likelihood (\ref{hstandlik}). The best fit value of the transition distance $d_{crit}$ is consistent with corresponding results of previous studies \cite{Alestas:2021nmi,Perivolaropoulos:2021bds,Perivolaropoulos:2022vql,Perivolaropoulos:2022khd}\footnote{See e.g. Figs. 8, 9 of Ref. \cite{Perivolaropoulos:2022khd}.} that have found hints for a transition of astrophysical properties including the parameters of the Tully-Fisher relation at a distance of about $20Mpc$. 

In Ref. \cite{Perivolaropoulos:2022khd} it was pointed out that if the SH0ES data are reanalyzed by allowing for a change of the SnIa absolute magnitude at $d_{crit}=50Mpc$ then the best fit value of the Hubble parameter shifts to a value almost identical with the inverse distance ladder best fit value albeit with significantly increased uncertainties. Motivated by this study we set $d_{crit}=50Mpc$ ($\mu_{crit}=33.5$) and minimize the generalized $\chi''^2(M_<,M_>,h,\Omega_{0m})$. We thus find the following best fit parameter values with the corresponding $1\sigma$ uncertainties
\ba
M_< =&-19.25\pm 0.03,\\
M_> =&-19.21\pm 0.05, \\
h =& 0.747 \pm 0.02, \\
\Omega_{0m} =& 0.33\pm 0.02, 
\ea
with minor change of the quality of fit since $\chi''^2_{min}=1522.28$ ($\Delta \chi''^2_{min}=-0.7$). Thus for $d_{crit}=50Mpc$ we find no hint of discrepancy between $M_<$ and $M_>$ and thus no inhomogeneity with respect to the SnIa intrinsic luminosities. In addition no significant change is observed in the best fit value of $h$ in contrast to the corresponding result for the SH0ES data analysis where the same degree of freedom induced a shift in the best fit value of $h$ to $h=0.67\pm 0.04$. This may be due to the small number of SnIa in Cepheid hosts for the $M_>$ bin (4 SnIa) combined with the much larger number of SnIa in the Hubble flow bin for the Pantheon+ sample and the much more extensive covariance matrix.

\begin{figure*}
\centering
\includegraphics[width =0.98 \textwidth]{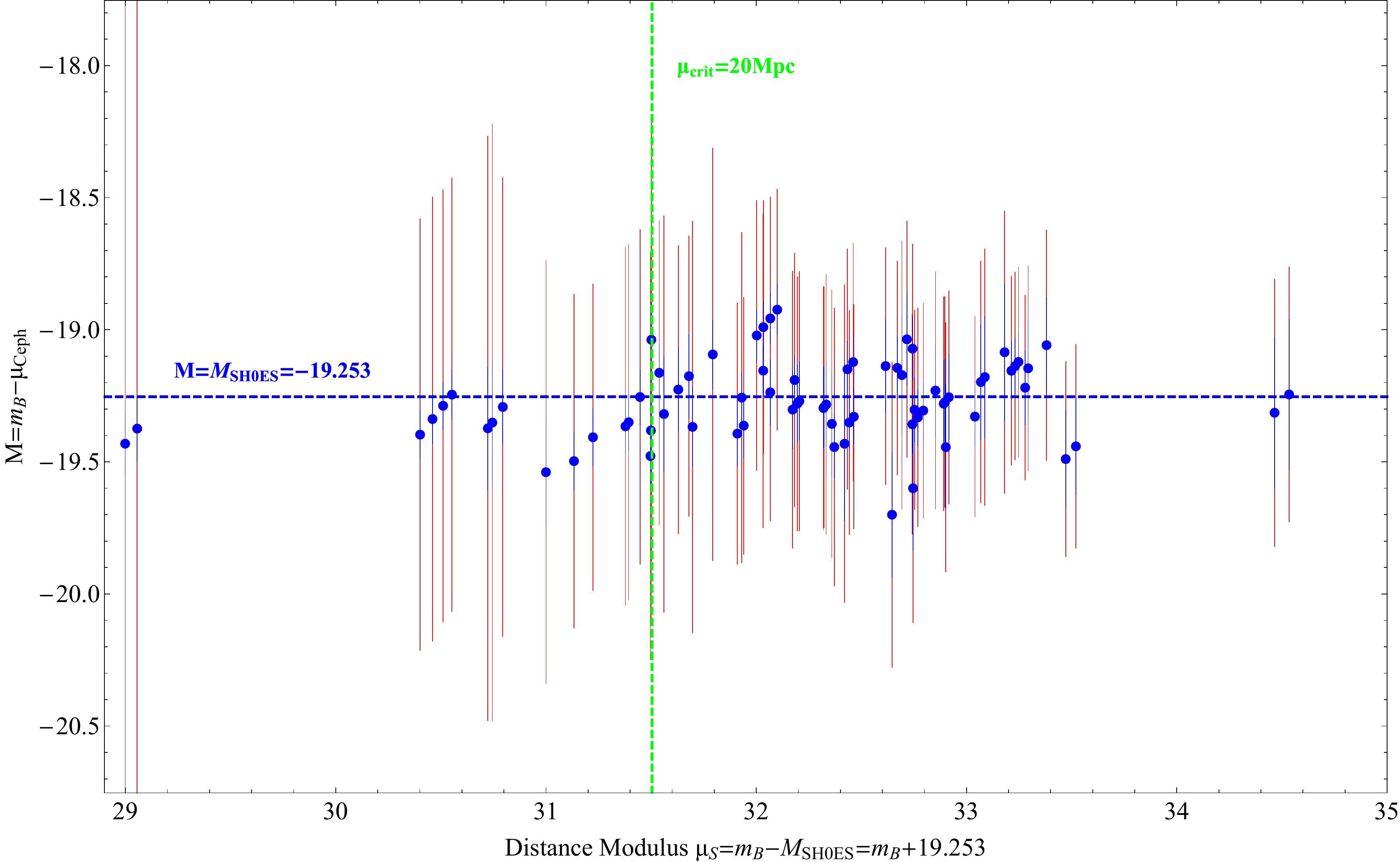}
\caption{The absolute magnitudes of SnIa residing in Cepheid hosts as obtained from the Pantheon+ data. The best fit SnIa standardized and corrected absolute magnitudes and uncertainties based on the SH0ES analysis is also shown (blue dashed line). The inflated uncertainties of the Pantheon+ data are due to redshift uncertainties due to peculiar velocity noise. Notice that all datapoints below the distance $d_{crit}=20Mpc$ are below the best fit $M_{SH0ES}$ line. } 
\label{fig3} 
\end{figure*}

\section{Statistical properties of SnIa intrinsic luminosities.}
\label{IV}
In this section we investigate the extend to which the discrepancy between the $M_<$ and $M_>$ parameters obtained in the previous section is due to a transition of the SnIa intrinsic luminosity or if it only due to the volumetric redshift scatter bias systematic. 
We thus focus on the particular subset of the Pantheon+ sample that corresponds to SnIa in Cepheid hosts and investigate the statistical properties of their individual absolute magnitudes. 
The measured absolute magnitude $M_i$ of individual SnIa in Cepheid hosts can be directly obtained from the Pantheon+ data as
\be
M_i=m_{Bi} - \mu_i^{{\rm Ceph}}
\label{Mi}
\ee
i.e. by subtracting column 13 from column 9 for those entries where column 14 is 1. The inflated uncertainty of each $M_i$ (red errorbar lines) is obtained from the corresponding entries of columns 10 and 13 only for plotting purposes as it will not be used in the statistical analysis of this section. This inflated uncertainty is not due to uncertainty of the SnIa luminosity but is due to the contribution of the peculiar velocity variation and its contribution to the redshift which should be taken into account in the Hubble diagram. It also is imposed to reduce the effect of the volumetric redshift scatter bias discussed in the previous section (for a more detailed discussion see Refs. \cite{Kenworthy:2022jdh,Brout:2022vxf}). 

%In the same plot we also show the pure SnIa luminosity errorbars (blue lines) as obtained from the SH0ES analysis (Table 6 of \cite{Riess:2021jrx}). These are the errorbars relevant for the SnIa absolute luminosisties as they do not include the irrelevant for our purposes of this section, redshift uncertainties from peculiar velocities. 

In Fig. \ref{fig3} we show a plot of the measured $M_i$ for 77 SnIa light curves in Cepheid hosts vs the SnIa distance moduli $\mu_{S,i}=m_{B,i}-M_{SH0ES}$  (column 9) as obtained from the measured apparent magnitudes and the best fit value of $M$ from SH0ES ($M_{SH0ES}=-19.253$, blue dashed line). We show both types of errorbars: The Pantheon+ errorbars  dominated by redshift uncertainties (red lines) and the SH0ES pure SnIa luminosity errorbars (blue lines) obtained using Table 6 of \cite{Riess:2021jrx}. 

As shown in Fig. \ref{fig3}, SnIa at distances $d<20Mpc$ appear to be systematically more luminus (lower $M_i$) than the rest of the SnIa since almost all are below the blue dashed line corresponding to $M_{SH0ES}$\footnote{A similar trend appears for the four furthest SnIa with $d>50Mpc$ even though this is much less significant statistically. This is the origin of the effects observed in \cite{Perivolaropoulos:2022khd}.}.

\begin{figure}
\centering
\includegraphics[width = \columnwidth]{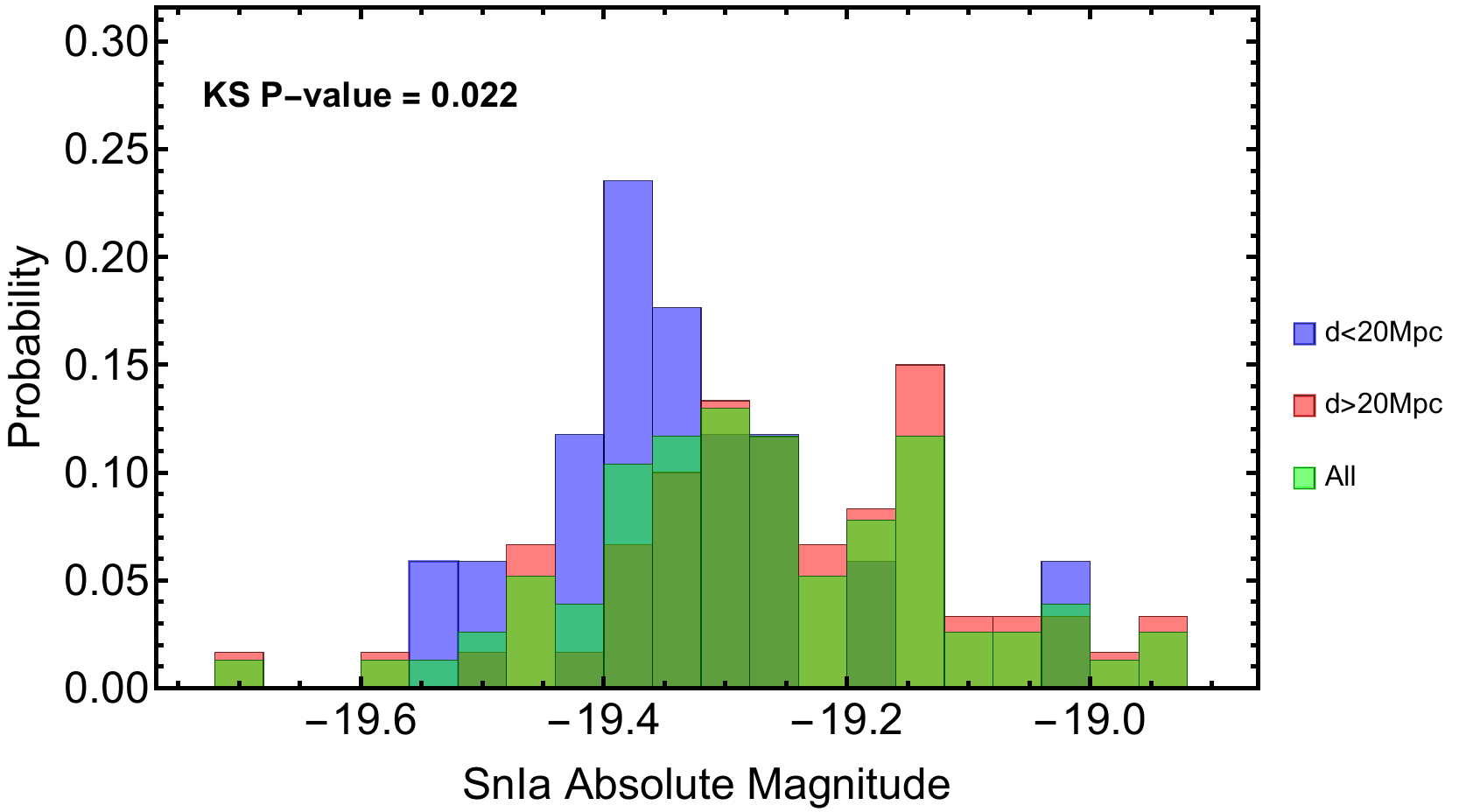}
\caption{The probability distribution histogram of the absolute magnitudes of each SnIa subsample (Nearby subsample $M_s^<$: dark blue columns, Distant subsample $M_s^>$: red columns, Full sample of SnIa in Cepheid hosts: green columns)} 
\label{fig4} 
\end{figure}

In order to compare the statistical properties of the $M_i$ subsample with $d<20Mpc$ ($M_s^<$) with the corresponding distant subsample with $d>20Mpc$ ($M_s^>$) we show in Fig. \ref{fig4} the probability distribution histogram of each subsample ($M_s^<$: dark blue columns, $M_s^>$: red columns, Full sample: green columns). As expected from Fig. \ref{fig3} the probability distributions for the two subsamples differ significantly with $M_s^<$ being significantly skewed towards lower $M$ values (brighter SnIa).

The statistical difference between the two subsamples may be quantified using a Kolmogorov-Smirnov test. Based on this test, the null hypothesis that the two subsamples have been drawn from the same probability distribution is rejected at the 2.5\% level. In fact the probability that the two subsamples have been drawn from the same probability distribution is 2.2\% (Kolmogorov-Smirnov P-value (KSPV))\footnote{The distance scale used for this estimate is based on the Cepheid distance modulus $\mu_{Ceph}$ (Column 13 of the Pantheon+ data). If we use the distance scale $\mu_S$ shown in Fig. \ref{fig3} (as done in all other parts of this analysis), the KSPV would be even smaller ($KSPV=0.001$).}  This result implies that a significant part of the $M_<-M_>$ discrepancy found in the previous section may be due to inhomogeneities in the SnIa corrected intrinsic luminosities of the Pantheon+ sample. This inhomogeneity could be due to either a large statistical fluctuation, or to  an unaccounted systematic effect or to a physics  transition that has occurred at a distance of about $20Mpc$ (about $70Myrs$ ago) \cite{Marra:2021fvf,Alestas:2020zol}. 

In the construction of the histogram of Fig. \ref{fig4} we have implicitly assumed that the 77 absolute magnitudes of SnIa light curves shown in Fig. \ref{fig3} are statistically independent. However, there are correlations among those light curves that refer to the same SnIa or to different SnIa in the same host. 

In order to eliminate these correlations at the expense of ignoring some useful information, we consider the merged data corresponding to the same host using a weighted average, based on the uncertainties of each one of the 77 absolute magnitudes of Fig. \ref{fig3}. This weighted, host based, merging can also be obtained using the data in Table 6 of the SH0ES analysis \cite{Riess:2021jrx}.

\begin{figure}
\centering
\includegraphics[width = \columnwidth]{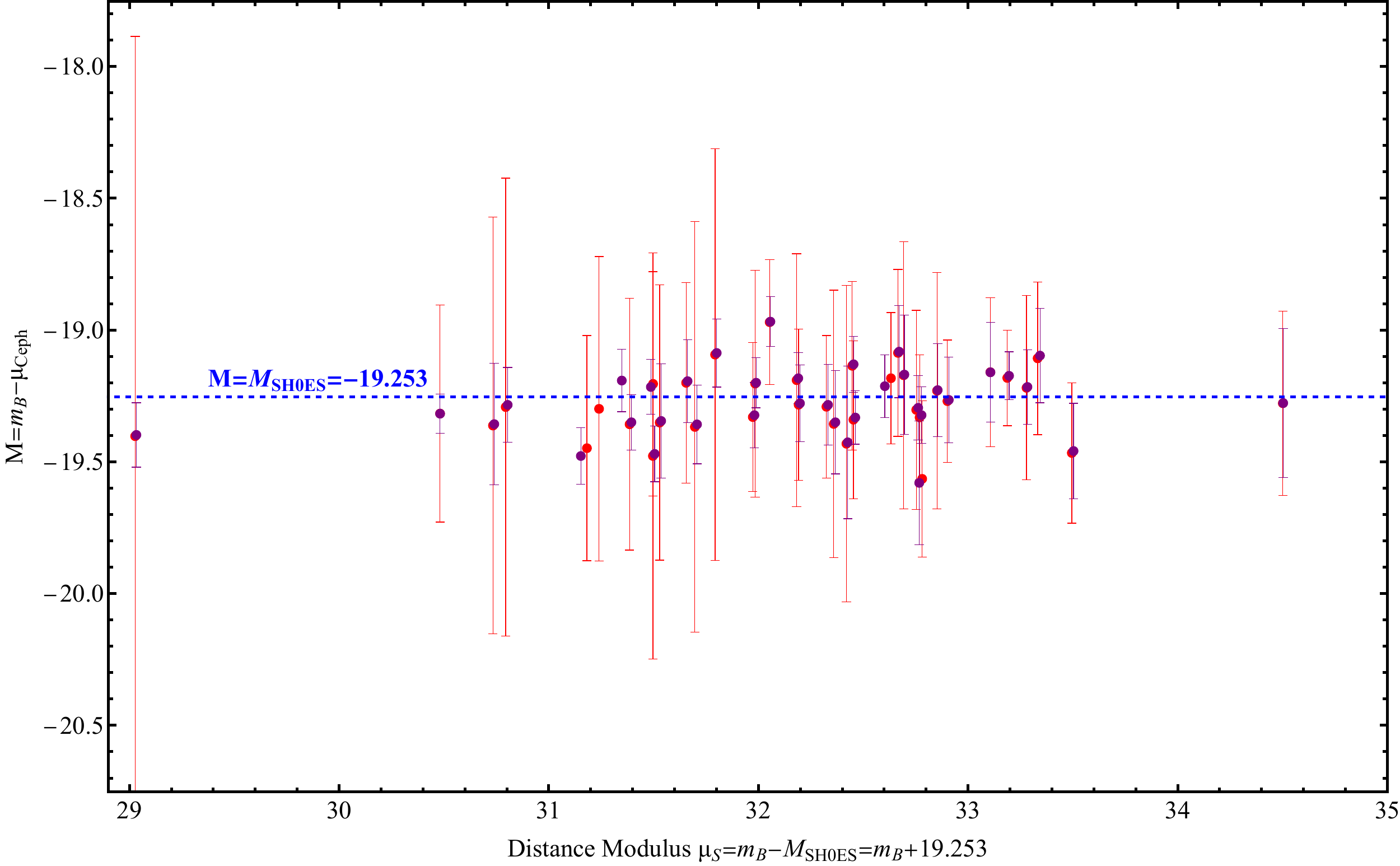}
\caption{The merged absolute magnitudes of SnIa+Cepheid hosts as obtained from the Pantheon+ data (red points) and from the SH0ES data (Table 6 of \cite{Riess:2021jrx}, purple points). The agreement is good but the points are not fully identical. The increased errorbars of the Pantheon+ data are due to the peculiar velocity uncertainties needed for the cosmological fits.} 
\label{fig5} 
\end{figure}

\begin{figure}
\centering
\includegraphics[width = \columnwidth]{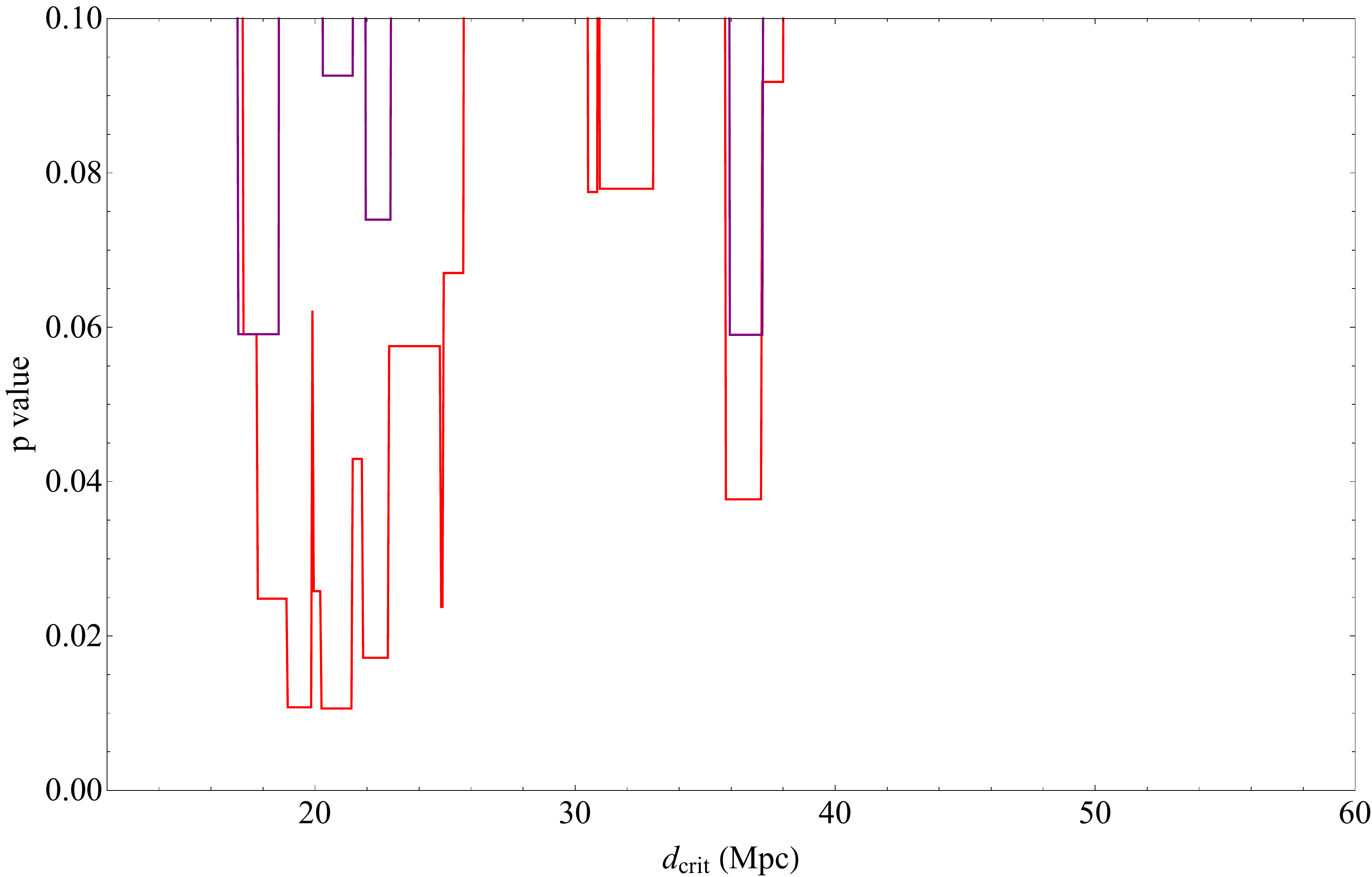}
\caption{The KSPV comparing the two distance bins of the data shown in Fig. \ref{fig5}  as a function of $d_{crit}$ that defines the two distance bins. The red line corresponds to the Pantheon+ merged data and the purple line corresponds to the SH0ES merged data.} 
\label{fig6}  
\end{figure}

%\section{Monte Carlo Analysis}
%\label{V}

We have thus reconstructed Fig. \ref{fig5} using weighted binning (merging) of the absolute magnitudes in each host.
The weighted binning in the $j^{th}$ Cepheid+SnIa host of the SnIa/light curve absolute magnitudes was implemented by considering the absolute magnitude of the $i^{th}$ SnIa/light curve in the $j^{th}$  host galaxy ($j=1-37$) and minimizing each $\chi^2(M_j)$ with respect to the $M_j$
\be
\chi^2(M_j)=\sum_{i=1}^{N_j}\frac{(M_i-M_j)^2}{\sigma_{M_i}^2}
\label{chiM}
\ee
where $N_j$ is the number of SnIa/light curves in the $j^{th}$ Cepheid host galaxy. The $1\sigma$ range $\delta M_j$ of each $M_j$  was obtained by solving $\Delta \chi^2= \chi^2(M) -\chi_{min}^2$ with $\Delta \chi^2=1$. This can be done analytically for the $j^{th}$ host \cite{Press2007}
leading to
\be
M_{j}=\frac{\sum_{i=1}^{N_j}M_i/\sigma_i^2}{\sum_{i=1}^{N_j} 1/\sigma_i^2}
\ee
\be
\sigma^2(M_{j})=\frac{1}{\sum_{i=1}^{N_j} 1/\sigma_i^2}
\ee

\begin{figure}
\centering
\includegraphics[width = \columnwidth]{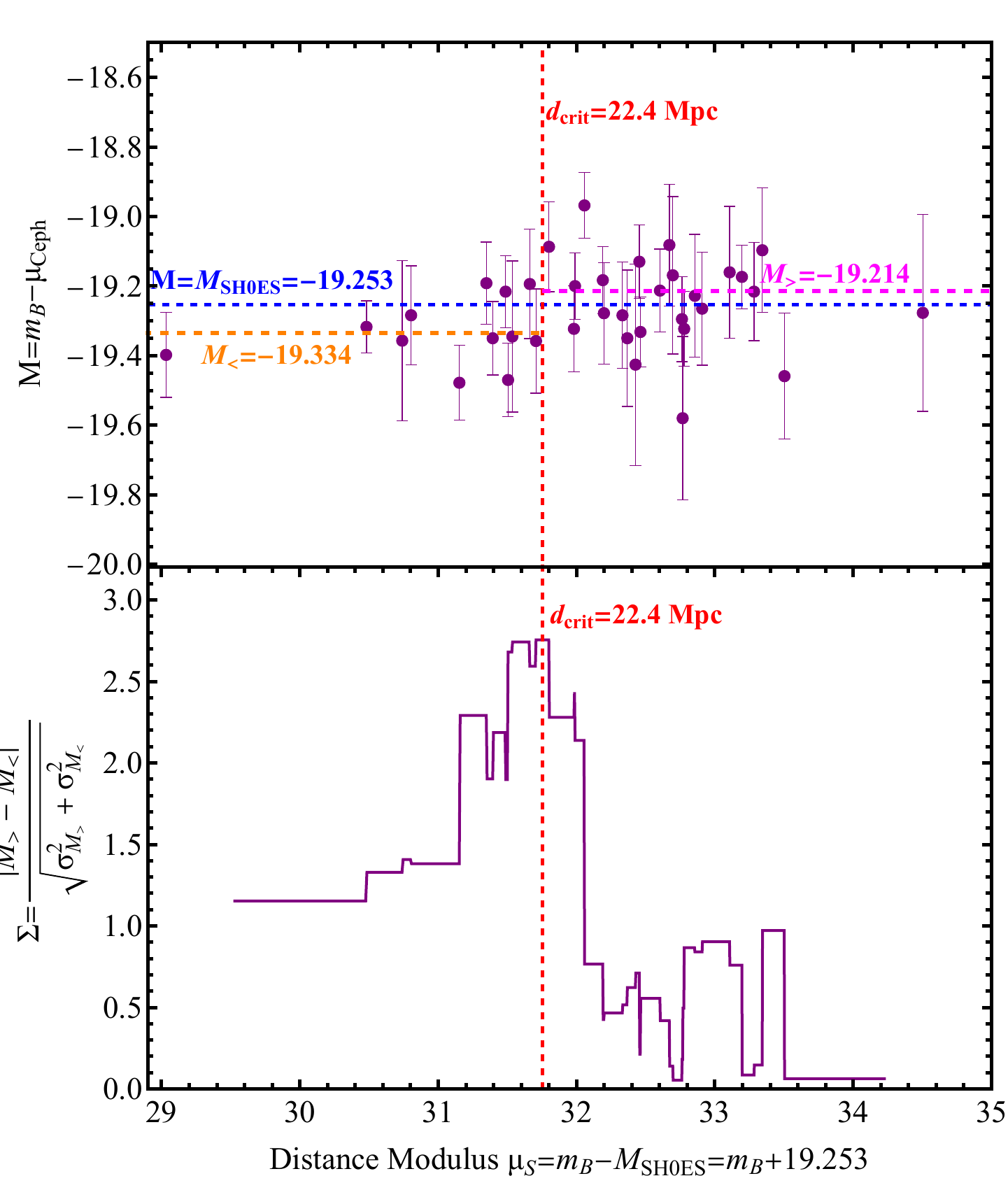}
\caption{{ \it Upper panel:} The 37 merged absolute magnitudes corresponding to the SnIa+Cepheid hosts obtained from Table \ref{tab:sh0es} as a function of the SH0ES distance modulus $\mu_S$. The distance $d_{crit}=22.4Mpc$ corresponding to the maximum transition significance $\Sigma_{max}$ is also indicated along with the corresponding values of $M_<$ and $M_>$ obtained from Eq. (\ref{msdef}). {\it Lower panel: } The significance $\Sigma$ of the $M_<-M_>$ transition as a function of the distance modulus $\mu_S^{crit}\equiv 5 log_{10}(d_{crit}/Mpc)+25$.}
\label{fig7}
\end{figure} 
where the index $i$ runs through the light curves or the SnIa of the $j^{th}$ host. Using this merging method on the 77 light curve data shown in Fig. \ref{fig3} or on the 42 SnIa shown in Table 6 of \cite{Riess:2021jrx}
we obtain the absolute magnitudes for each host shown in Fig. \ref{fig5} obtained with the Pantheon+ data of Fig. \ref{fig3} (red points) and from the SH0ES data (purple points, obtained using the data and uncertainties of Table 6 of \cite{Riess:2021jrx}). The agreement between the two sources is very good with minor differences in a very small number of datapoints.

\begin{figure}
\centering
\includegraphics[width = \columnwidth]{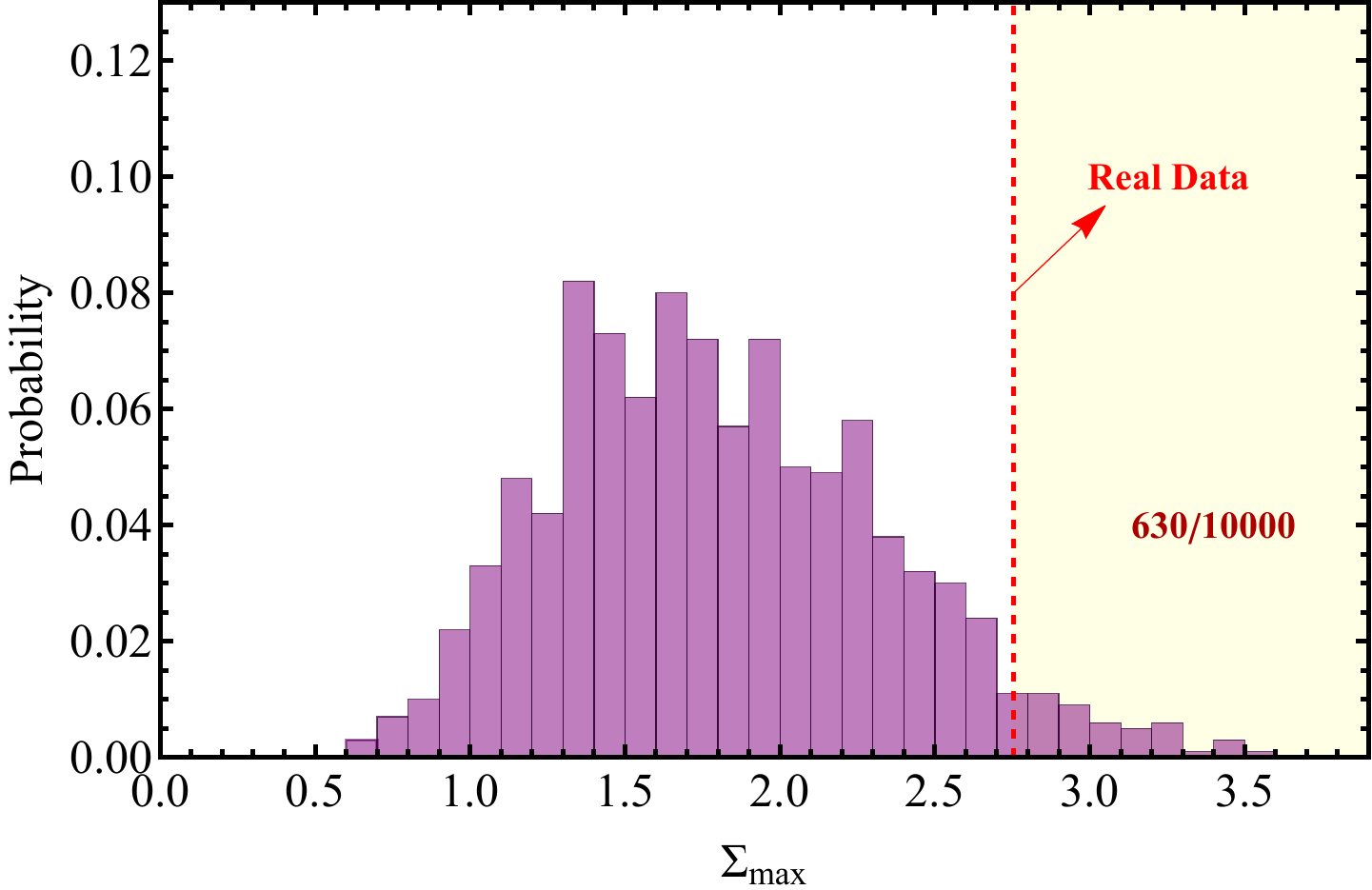}
\caption{The probability histogram for the maximum transition significance $\Sigma_{max}$ obtained from homogeneous simulated absolute magnitude data samples corresponding to the real data of Fig. \ref{fig7} (upper panel). The  $\Sigma_{max}$ value of the real data is also shown. }
\label{fig8}
\end{figure} 

We now split the data in low and high distance bins split at a distance $d_{crit}$ and compare the two bins to find the KSPV for each pair indicating the probability that the SnIa absolute magnitudes of each bin have been drown from the same probability distribution. The resulting KSPV as a function of the splitting distance $d_{crit}$ is shown in Fig. \ref{fig6}. The KSPV drops down to 0.01 for $d_{crit}\simeq 20Mpc$ for the Pantheon+ absolute magnitudes (red line) and down to 0.06 for the SH0ES absolute magnitudes of Fig. \ref{fig5}.

As an additional test of the homogeneity of the SnIa absolute magnitudes we investigate the statistical properties of the significance $\Sigma$ of the $M_<-M_>$ transition.
The SH0ES merged host absolute magnitudes along with their uncertainties are plotted in the upper panel of Fig. \ref{fig7} in terms of the merged SH0ES distance moduli defined as $\mu_{Si} \equiv m_{Bi}-M_{SH0ES}\equiv m_{Bi}+19.253$ (upper panel). We then split these data in two distance bins according to a critical distance $d_{crit}$ or a critical distance modulus $\mu_{crit}=5 log(d_{crit}/Mpc)+25$. For the low distance bin $\mu<\mu_{crit}$ we obtain the weighted mean absolute magnitude $M_<$ and its uncertainty $\sigma^2(M_{<})$ as
\be
M_{<}=\frac{\sum_{i=1}^{N_k}M_i/\sigma_i^2}{\sum_{i=1}^{N_k} 1/\sigma_i^2}
\label{msdef}
\ee
\be
\sigma^2(M_{<})=\frac{1}{\sum_{i=1}^{N_k} 1/\sigma_i^2}
\ee
where the index $i$ runs over the $N_k$ low distance bin hosts. Similarly we also obtain the high distance bin absolute magnitude $M_>$ and its uncertainty  $\sigma^2(M_{>})$. Then for each value of $\mu_{crit}$ we find the transition significance defined as 
\be 
\Sigma(\mu_{crit})\equiv \frac{|M_>-M_<|}{\sqrt{\sigma_{M_>}^2+\sigma_{M_<}^2}}
\label{sigmadef}
\ee
and plot $\Sigma(\mu_{crit})$ in the lower panel of Fig. \ref{fig7} in terms of $\mu_S^{crit}$. The maximum significance $\Sigma_{max}$ is obtained for $d_{crit}=22.4 Mpc$ as $\Sigma_{max}=2.75$ (lower panel of Fig. \ref{fig7} ) with
\ba
M_<(d_{crit}&=&22.4)=-19.33\pm 0.04\\
M_>(d_{crit}&=&22.4)=-19.21\pm 0.03
\ea
These values are shown as dashed lines in the upper panel of Fig.  \ref{fig7}  based on the 37 host merged SH0ES datapoints shown in Table \ref{tab:sh0es} which in turn is based on the 42 SnIa shown in Table 6 of \cite{Riess:2021jrx}.

Next we address the question: How often would this value of $\Sigma_{max}$ or larger  be obtained by chance in the context of Monte Carlo simulations generated under the assumption of homogeneous data with Gaussian variation around the best fit value of $M=M_{SH0ES}=-19.253$ with the same uncertainties as the real data? Thus  we generate 10000 such homogeneous simulated samples and vary $d_{crit}$ in each of them (with step $\Delta d_{crit}=0.2$) to determine the maximum transition significance $\Sigma_{max}(\mu_{crit})$ in each simulated sample.  We find  about that 630 such random samples (6.3\% probability) have $\Sigma_{max}$ larger or equal to the $\Sigma_{max}$ of the real data. We conclude that the probability that a real transition signal is hidden in the SnIa absolute magnitudes of the data is about $94\%$. This probability would be somewhat lower if we had used the $\mu_{Ceph}$ (column 1 of Table \ref{tab:sh0es}) distance scale instead of the $\mu_S$ (column 3 of Table \ref{tab:sh0es}) distance scale in defining the $<$ and $>$ distance bins.

\section{Conclusion-Discussion}
\label{V}
We have tested the internal consistency of the Pantheon+ sample with respect to the absolute magnitude parameter $M$. We have allowed for a change of $M$ at some transition distance $d_{crit}$  from $M_<$ at low distances (late times) to $M_>$ at high distances (early times). For $d_{crit}= 19.95 Mpc$, we found that such a change is favored by the Pantheon+ data leading to a reduction of $\chi^2$ by $\Delta \chi^2_{min}=-19.6$ in the context of a \lcdm cosmological background. This  corresponds to reduction of the Akaike Information Criterion (AIC) by $\Delta AIC=-15.5$ (for two additional parameters). Such a reduction provides strong evidence that the model where a change of $M$ is allowed, is very strongly preferred by the Pantheon+ data over the baseline model with a single value for $M$. This conclusion is further amplified by the fact that the best fit value of $M_<$ is in discrepancy with the best fit value of $M_>$ at a level of more than $3\sigma$.

This effect is due at least partly, to the volumetric redshift scatter bias which is well known to exist in the data\cite{Kenworthy:2022jdh,Brout:2022vxf} for Hublle diagram points with $z<0.01$. We have tested the hypothesis that the $M_<-M_>$ discrepancy is due solely to this bias. We thus removed all Hubble diagram redshift datapoints with $z<0.01$ from the Pantheon+ sample. We showed that the corresponding improvement of the fit reduces from $\Delta \chi^2 = -19.6$ to $\Delta \chi^2 = -7.5$. For the assumed two additional parameters this corresponds to $\Delta AIC=-3.5$ which indicates mild preference for an intrinsic SnIa absolute luminosity transition at $d_{crit}\simeq 20 Mpc$. The tension between the best fit parameters $M_<$ and $M_>$ also reduces to a little less than $2\sigma$. 

The probability of existence of an intrinsic transition of SnIa intrinsic luminosity in the Pantheon+ and SH0ES data was also estimated using Monte Carlo simulations to be about $94\%$.  In addition,  a Kolmogorov-Smirnov test has indicated that the probability that the  absolute magnitudes $M_i$ of SnIa in Cepheid hosts at distances $d<22.4Mpc$ are drawn from the same distribution as the $M_i$ of SnIa at hosts with $d>22.4Mpc$ is less than 1.5\%. This level of significance however, does not necessarily correspond to a transition of SnIa absolute luminosity and also is consistent with about 7\% of homogeneous Monte Carlo data samples. 

There are three main implications of our results:

{\bf New Likelihood Model:} The likelihood (\ref{eq:qdprime}) provides a much better fit to the Pantheon+ data than the standard likelihood (\ref{eq:qprimedef}) used in all current analyses of the Pantheon+ data. An important reason for this is the presence of the volumetric redshift scatter bias which is well fit by the new parameter $M_<$ in the (\ref{eq:qdprime}) likelihood model. The use of this likelihood instead of the standard one of Brout et. al. \cite{Brout:2022vxf} does not affect the best fit value  of the cosmological parameter $\Omega_{0m}$ as shown by comparing Eqs. (\ref{omstand}) and (\ref{omnlm}) but it does mildly affect the best fit value of the Hubble parameter $h$ raising it from $h=0.734 \pm 0.1$ to $h=0.749\pm 0.1$ (see Eqs. (\ref{hstandlik}) and (\ref{hnlm})). It may therefore affect other cosmological parameters of dynamical dark energy that are sensitive to low redshift cosmic dynamics. Hints for such an effect have been recently pointed out by \citep{Pasten:2023rpc} where it was shown that the deceleration parameter is affected by the presence of the $z<0.01$ data of Pantheon+. For such cosmological models it may be more appropriate to use the likelihood (\ref{eq:qdprime}) (or a generalized version of it) to fit the corresponding cosmological parameters. Thus a more detailed study of the effects of the new likelihood model on the best fit values of cosmological parameters would be an interesting extension of the present analysis.  

{\bf Mild Evidence for Change of SnIa luminosity:} The mild evidence for a transition of the SnIa absolute luminosity at a distance of about $20Mpc$ requires further testing. If this possible SnIa luminosity inhomogeneity in the Pantheon+ and SH0ES samples is due to a physics transition \cite{Alestas:2022xxm,Mortonson:2009qq,Alestas:2021luu,Benevento:2020fev,DiValentino:2017rcr,Alestas:2020zol,Banihashemi:2018oxo,DiValentino:2019exe,Keeley:2019esp,Caprini:2019egz,Farhang:2020sij,Sato:1980yn,Patwardhan:2014iha} (e.g. gravitational transition \cite{Caldwell:2005xb,Marra:2021fvf,Khosravi:2017hfi,Perivolaropoulos:2022txg,pres1,pres2}), it may have implications for other calibration parameters like the color and stretch parameters. It would therefore be of interest to extend the present analysis in such directions by testing the homogeneity of the Pantheon+ sample with respect to possible differences of the best fit values of such parameters when these are allowed to change among different subsamples of the full Pantheon+ sample. Hints for such inhomogeneities from the first Pantheon sample have been already reported \cite{Wojtak:2022bct} with respect to the color and stretch parameters. 

{\bf Implications for the Hubble tension:} The best fit value $M_< =-19.33 \pm 0.04$ \footnote{In the absence of the volumetric redshift bias (no redshift points for $z<0.01$).} of the low distance SnIa absolute magnitude is consistent with the inverse distance ladder best fit value $M=-19.4\pm 0.027$ \cite{Marra:2021fvf}. This coincidence may have implications for the Hubble tension because if for some physical reason the best fit value of $M_>$  is not reliable and the true value of the SnIa luminosity is that implied by the best fit $M_<$ then the Hubble tension would be eliminated. In fact this observation may be related to the reduced tension found when calibrators at lower distances compared to Cepheids (like TRGB \cite{Freedman:2019jwv}) are used to calibrate SnIa.  Thus the dependence the inferred SnIA absolute magnitude $M_i$, obtained from other SnIa calibrators like TRGB \cite{Freedman:2019jwv}, on distance, should also be investigated to identify similar possible distance dependence on the calibration parameters. 

As pointed out in previous studies, possible evolution of SnIa properties either in the form of a transition or in the form of smooth evolution could significantly modify the values of cosmological parameters as obtained from data in different redshift bins \cite{Colgain:2022rxy,Colgain:2022nlb,Kazantzidis:2020xta,Kazantzidis:2020tko}. Even if there is no real evolution of SnIa physical properties and the observed effect is purely due to volumetric redshift scatter bias or another systematic, its proper consideration in the likelihood through the likelihood model of Eq. (\ref{eq:qdprime}) may play a role in the more accurate determination of cosmological parameters in the context of specific dynamical dark energy models with low $z$ evolution. The investigation of this effect may be implemented by comparing the values of a wide range of cosmological dynamical dark energy models using the new likelihood (\ref{eq:qdprime}) as an alternative to the standard Pantheon+ likelihood model (\ref{eq:qprimedef}).

\begin{widetext}
\begin{table*}    
\caption{The SnIa+Cepheid host merged SH0ES data used for the construction of Figs. \ref{fig5} and \ref{fig7}. The line corresponds to the $d_{crit}=22.4Mpc$ as obtained from the $\mu_S$ column.}
\label{tab:sh0es} 
\vspace{2mm}

\begin{tabular}{p{1 cm} p{1.5cm} p{1.5cm} p{1.5cm}p{1.5cm} p{1.5cm} p{1.5cm} p{1.5cm} p{1.5cm} p{1.5cm}} 
\hline
\hline
   & &&  &  &  & &  &  &  \\ 
 N &Host &$\;\mu_{Ceph}$& $\;\;\;\sigma$& $\;\;\mu_S^a$&$\;\;\;\sigma$ &\;\;\;M&$\;\;\;\sigma$& $\;\;m_B$ &$\;\;\;\sigma$ \\
  & &[mag]& [mag] & [mag] & [mag] &[mag] & [mag] & [mag] & [mag]  \\
  & &&  &  &  & &  &  &  \\ 
\hline
\hline
1 &M101 &29.178 &0.041 &29.033 &0.119 &-19.398 &0.122 &9.78 &0.115\\
2 &N5643 &30.546 &0.052 &30.482 &0.062 &-19.317 &0.075 &11.229 &0.054\\
3 &N4424 &30.844 &0.128 &30.740 &0.194 &-19.357 &0.231 &11.487 &0.192\\
4 &N4536 &30.835 &0.05 &30.804 &0.136 &-19.284 &0.142 &11.551 &0.133\\
5 &N1365 &31.378 &0.056 &31.153 &0.097 &-19.478 &0.108 &11.9 &0.092\\
6 &N1448 &31.287 &0.037 &31.348 &0.116 &-19.192 &0.118 &12.095 &0.112\\
7 &N1559 &31.491 &0.061 &31.394 &0.091 &-19.35 &0.105 &12.141 &0.086\\
8 &N2442 &31.45 &0.064 &31.487 &0.087 &-19.216 &0.104 &12.234 &0.082\\
9 &N3982 &31.722 &0.071 &31.505 &0.084 &-19.47 &0.105 &12.252 &0.078\\
10 &N7250 &31.628 &0.125 &31.536 &0.181 &-19.345 &0.218 &12.283 &0.178\\
11 &N4038 &31.603 &0.116 &31.662 &0.11 &-19.194 &0.157 &12.409 &0.106\\
12 &N4639 &31.812 &0.084 &31.707 &0.128 &-19.358 &0.15 &12.454 &0.124\\
 & &&  &  &  & &  &  &  \\ 
\hline
  & &&  &  &  & &  &  &  \\ 
13 &N3972 &31.635 &0.089 &31.801 &0.099 &-19.087 &0.129 &12.548 &0.094\\
14 &N2525 &32.051 &0.099 &31.981 &0.08 &-19.323 &0.124 &12.728 &0.074\\
15 &N3447 &31.936 &0.034 &31.989 &0.094 &-19.2 &0.095 &12.736 &0.089\\
16 &N5584 &31.772 &0.052 &32.057 &0.085 &-18.968 &0.095 &12.804 &0.079\\
17 &N3370 &32.12 &0.051 &32.19 &0.087 &-19.183 &0.097 &12.937 &0.082\\
18 &N5861 &32.223 &0.099 &32.198 &0.111 &-19.278 &0.146 &12.945 &0.107\\
19 &N5917 &32.363 &0.12 &32.332 &0.1 &-19.284 &0.153 &13.079 &0.095\\
20 &N3021 &32.464 &0.158 &32.367 &0.12 &-19.35 &0.196 &13.114 &0.116\\
21 &N4680 &32.599 &0.205 &32.426 &0.207 &-19.426 &0.29 &13.173 &0.205\\
22 &N3254 &32.331 &0.076 &32.454 &0.08 &-19.13 &0.106 &13.201 &0.074\\
23 &N1309 &32.541 &0.059 &32.462 &0.087 &-19.332 &0.101 &13.209 &0.082\\
24 &N1015 &32.563 &0.074 &32.603 &0.099 &-19.213 &0.12 &13.35 &0.094\\
25 &N7541 &32.5 &0.119 &32.671 &0.131 &-19.082 &0.175 &13.418 &0.128\\
26 &N2608 &32.612 &0.154 &32.696 &0.169 &-19.169 &0.226 &13.443 &0.166\\
27 &N3583 &32.804 &0.08 &32.762 &0.098 &-19.295 &0.123 &13.509 &0.093\\
28 &N5728 &33.094 &0.205&32.767 &0.119 &-19.580 &0.235 &13.514 &0.115\\
29 &U9391 &32.848 &0.067 &32.778 &0.089 &-19.323 &0.107 &13.525 &0.084\\
30 &N0691 &32.83 &0.109 &32.855 &0.142 &-19.228 &0.177 &13.602 &0.139\\
31 &M1337 &32.92 &0.123 &32.908 &0.11 &-19.265 &0.162 &13.655 &0.106\\
32 &N3147 &33.014 &0.165 &33.107 &0.098 &-19.16 &0.19 &13.854 &0.093\\
33 &N5468 &33.116 &0.074 &33.195 &0.061 &-19.174 &0.091 &13.942 &0.054\\
34 &N7329 &33.246 &0.117 &33.283 &0.085 &-19.216 &0.141 &14.030 &0.079\\
35 &N7678 &33.187 &0.153 &33.343 &0.098 &-19.097 &0.179 &14.09 &0.093\\
36 &N0976 &33.709 &0.149 &33.503 &0.107 &-19.459 &0.181 &14.25 &0.103\\
37 &N0105 &34.527 &0.25 &34.503 &0.136 &-19.277 &0.283 &15.25 &0.133\\
\hline
\hline
\end{tabular} 

\noindent{\footnotesize{Note: (a) The ranking order of the host galaxies in the table was made with increasing distance modulus $\mu_S$ (where $\mu_S=m_B-M_{SH0ES}=m_B+19.253$)}}
\end{table*}

\end{widetext}

\section*{Numerical Analysis Files}

The numerical files for the reproduction of the figures can be found  \href{https://github.com/leandros11/pantheonplus1}{this Github repository under the MIT license.} \\

\section*{Acknowledgments}
Special thanks are due to Adam Riess for extensive discussions and comments that significantly improved our analysis and its interpretation. We also thank Eoin Colgain, Dan Scolnic and Dillon Brout for useful comments. This article is based upon work from COST Action CA21136 - Addressing observational tensions in cosmology with systematics and fundamental physics (CosmoVerse), supported by COST (European Cooperation in Science and Technology). This project was also supported by the Hellenic Foundation for Research and Innovation (H.F.R.I.), under the "First call for H.F.R.I. Research Projects to support Faculty members and Researchers and the procurement of high-cost research
equipment Grant" (Project Number: 789).\\

\raggedleft
\bibliography{Bibliography}

\end{document}